\documentstyle[preprint,aps]{revtex}
\tighten
\begin{document}

\draft

\newcommand{\lsim}{{ _< \atop ^\sim}}
\newcommand{\gsim}{{ _> \atop ^\sim}}

\title{Temperature and magnetic field dependence of the lattice
constant in spin--Peierls cuprate $\bf CuGeO_3$ studied by capacitance
dilatometry in fields up to 16 Tesla}

\author{T. Lorenz, U. Ammerahl, T. Auweiler, and B. B\"uchner}
\address{II. Physikalisches Institut, Universit\"at zu K\"oln,
         Z\"ulpicher Str. 77, 50937 K\"oln, Germany}

\author{A.Revcolevschi and G. Dhalenne}
\address{Laboratoire de Chimie des Solides, Universit\'e Paris-Sud,
         91405 Orsay C\'edex, France}

\date{\today}

\maketitle

\begin{abstract}
We present high resolution measurements of the thermal expansion coefficient and
the magnetostriction along the $a$--axis of $\rm CuGeO_3$ in magnetic fields up
to 16 Tesla. From the pronounced anomalies of the lattice constant $a$ occurring
for both temperature and field induced phase transitions clear structural
differences between the uniform, dimerized, and incommensurate phases are
established. A precise field temperature phase diagram is derived and compared
in detail with existing theories. Although there is a fair agreement with the
calculations within the Cross Fisher theory, some significant and systematic
deviations are present. In addition, our data yield a high resolution
measurement
of the field and temperature dependence of the spontaneous strain scaling with
the spin--Peierls order parameter. Both the zero temperature values as well as
the critical behavior of the order parameter are nearly field independent in the
dimerized phase. A spontaneous strain is also found in the incommensurate high
field phase, which is significantly smaller and shows a different critical
behavior than that in the low field phase.
The analysis of the temperature dependence
of the spontaneous strain yields a pronounced field dependence
within the dimerized phase, whereas the temperature dependence
of the incommensurate lattice modulation compares well with that of the
dimerization in zero magnetic field.
\end{abstract}


\section{Introduction}

Since the discovery of a spin--Peierls transition in the inorganic cuprate
$\rm CuGeO_3$ by Hase et
al.~\cite{hase93a} this unusual magnetoelastic transition
occurring in quasi one--dimensional antiferromagnetic insulators
has again attracted considerable attention. Compared to the
well known organic spin--Peierls systems~\cite{reviewbray} the structure of
$\rm CuGeO_3$ is rather simple. This fact and the possibility
of growing large high quality single crystals allows for
a much better study of the spin--Peierls phenomena in $\rm CuGeO_3$ than in the
organic compounds.

Most properties expected from the well developed theory of the spin--Peierls
transition, e.g. the Cross Fisher (CF) theory~\cite{cross79a,cross79b},
are observed in $\rm CuGeO_3$. For example elastic
neutron and x--ray scattering
show a doubling of the orthorhombic unit cell
below the transition temperature $\rm T_{SP} \simeq 14.3K$
leading to two non--equivalent Cu sites in the magnetic chains.
This lattice distortion leads to alternating Cu--O--Cu superexchange
paths, i.e. alternating intrachain magnetic exchange constants.
Also in agreement with the CF theory
a gap in the magnetic excitations is observed~\cite{nishi94b},
which scales with the structural order parameter, i.e the dimerization.
Whereas these properties of the dimerized (D) phase of $\rm CuGeO_3$
seem to be well represented by a model
of  spin--1/2 Heisenberg chains with a spin--Peierls transition,
some significant deviations from this most simple treatment
are present in the uniform (U) phase, i.e. for $\rm T > T_{SP}$.
Most strikingly the magnetic susceptibility in the U phase of $\rm CuGeO_3$
disagrees with the temperature dependence calculated for
one--dimensional spin--1/2 Heisenberg
chains~\cite{hase93a,castilla95,riera95}.
There is evidence that the corrections which are
necessary to explain the magnetism in the U phase
also influence the spin--Peierls transition.
It is argued for example that due to a frustration
of the magnetic exchange in the quasi one--dimensional
chains~\cite{castilla95,buechner96a}
$\rm T_{SP}$ is strongly enhanced in $\rm CuGeO_3$.

The influence of a magnetic field represents
a further characteristic feature of the spin--Peierls transition, which can
be directly compared to the different theoretical predictions.
Due to the additional Zeeman term in the Hamiltonian
the nonmagnetic dimerized phase is destabilized when applying a magnetic field.
Consequently an additional phase with a finite susceptibility
occurs at high magnetic fields, in $\rm CuGeO_3$ for $H \gsim 12$ Tesla.
An incommensurate lattice modulation
has been predicted theoretically for this I phase and
was recently observed by x--ray diffraction~\cite{kiryukhin95b}.
However, the knowledge about this phase is still very limited,
e.g. the spatial character of the incommensurate lattice modulation --
domain walls or sinusoidal modulation -- is still a subject of debate.

The theoretical H--T phase diagrams
which have been calculated with different treatments
of spin 1/2 Heisenberg (or XY) chains differ significantly.
A detailed experimental determination of the phase diagram in $\rm CuGeO_3$
therefore allows for a test of these theories and future
descriptions incorporating e.g. a frustration of the magnetic exchange.
The theories yield different predictions for the
positions of the three phase boundaries -- U/D, U/I, and D/I --
in reduced field and temperature scales.
Moreover, the CF theory
and the earlier theory by Bulaevskii et al.~\cite{bulaevskii78} predict
a discontinuous first order D/I transition, whereas a continuous
transition is obtained within the soliton picture~\cite{fujita84}.

So far the H--T phase diagram of $\rm CuGeO_3$ has been mainly studied
via measurements of the magnetization.
The phase boundaries roughly agree with the predictions of the
CF theory~\cite{hase93c},
though when analyzing the data in detail some deviations
seem to be present. Up to now there is only
little information on the thermodynamic and structural
properties in high magnetic fields, particularly with regard to the
incommensurate phase.

In this paper we will present a detailed study of the
coefficient of the thermal expansion $\rm \alpha$
along the $a$--axis in magnetic fields up to 16 Tesla~\cite{otherdir}.
As shown in Ref.~\cite{winkelmann95}
there is a very large anomaly of $\rm \alpha$
at the U/D transition. Therefore, a very detailed investigation of the
H--T phase diagram is possible via measurements of $\rm \alpha$ in
external fields. Moreover, $\rm \alpha$ represents a
thermodynamic property similar to the specific heat. Therefore it allows
to study the nature of the different transitions
and the critical behavior. Furthermore,
$\rm \alpha$ is per definition a structural property,
i.e. the temperature dependence of the relative lattice constant.
Because of that the giant anomalies of $\rm \alpha$ at $\rm T_{SP}$
measure the development of spontaneous strains ($\rm \epsilon$) at
the U/D transition. The spontaneous strains
of the D phase in $\rm CuGeO_3$ are surprisingly rather large
($\rm \simeq 10^{-5}$)
and have been observed not only by high resolution capacitance methods
but also by diffraction techniques~\cite{lorenzo94,harris94,fujita95c}.

A comparison with neutron diffraction data reveals that
the spontaneous strains are proportional to the squared
order parameter of the spin--Peierls transition.
Thus, the high resolution measurements of $\rm \alpha$
yield a precise measurement of the order parameter in the D--phase
as a function of both temperature and magnetic field.
In addition, we will show that
reduced but still large spontaneous strains are present
in the I phase, and thus the temperature
dependence of the incommensurate lattice modulation can also
be determined from our data.

The thermal expansion measurements performed
at fixed fields yield information on the phase transitions
occurring as a function of temperature. In order to allow for the
discussion of a complete H--T phase diagram,
we will also present some measurements of the
magnetostriction~\cite{mehrkommt},
i.e. the changes of the lattice constant as a function of a magnetic
field at a fixed temperature.
The magnetostriction shows anomalies at field driven phase transitions
and thus it is possible to determine the D/I phase boundary
which is nearly parallel to the temperature axis in the
H--T phase diagram and therefore difficult to analyze by
thermal expansion measurements.

The paper is organized as follows. After a short description
of the capacitance dilatometer we will give an overview of
the experimental observations in Section~\ref{RESULTS}.
The pronounced differences between the lattice constants
of the three phases of $\rm CuGeO_3$ are already visible
in these raw data. The H--T phase diagram
obtained from our measurements
is presented and discussed in section~\ref{PHADI}.
The following section~\ref{FIELDDEP}
''{\em Field dependence of the thermal expansion and the spin--Peierls
order parameter}'' is divided into five parts.
Before discussing the
influence of the magnetic field on the lattice constant in detail
we will shortly repeat some results extractable
from $\rm \alpha$ in zero magnetic field.
Then we are going to discuss the reduction of the
anomaly of $\rm \alpha$ at $\rm T_{SP}$ with increasing field
by considering its
relationship to the anomaly of the specific heat.
The next part deals with the critical behavior of the order parameter,
which differs for the U/D and U/I transitions, respectively.
Hereafter the magnetic field dependence of the dimerization
at low temperatures, which is found to be extremely weak
in the D phase, and the pronounced
discontinuous changes of the spontaneous strains
at the field driven D/I transition are discussed.
Finally in the last part of section~\ref{FIELDDEP} the temperature
dependencies of the order parameter  well below $\rm T_{SP}$ are analyzed.
We find a remarkably strong field dependence
within the D phase, whereas the temperature
dependence of the incommensurate modulation compares well with that
of the dimerization in zero magnetic field.

\section{Experimental}

The single crystal of $\rm CuGeO_3$ used for the present
study was cut from a large crystal
(80 mm along the $a$--axis) grown by
a floating zone technique~\cite{revcolevschi93}.
The sample is of cylindrical shape with dimensions
of about 6x5x8.3 $\rm mm^3$ for the $a$--,$b$--, and $c$--axis,
respectively.
Various experimental investigations on
samples prepared in this way have already been published.
Some details of structural and magnetic
properties of these crystals are described e.g.
in Ref.~\cite{pouget94}.

The measurements were carried out with a new capacitance dilatometer.
It was originally designed to allow for measurements of the
coefficient of the thermal expansion
$\alpha \equiv {1 \over L } \cdot {\partial L \over \partial T}$
($L$: length of the sample) in fixed external fields only.
Besides, it enables to measure the magnetostriction,
i.e. the field induced length changes
${\Delta L(H) \over L}$ at fixed temperatures.
Both types of measurement can be performed during a single run, i.e.
for exactly the same orientation of the crystal in the dilatometer.
During the measurements the dilatometer is mounted into an evacuated
stainless steal tube, which fits into the 40 mm bore
of a superconducting 16 Tesla magnet.
Although the new dilatometer is based on our
conventional one described in Ref.~\cite{pott83}
some significant differences exist.
The main difference is a thermal decoupling between the sample and
the plate capacitor, i.e. the temperature of the sample is
changed, but the capacitor is thermally coupled to
the liquid He bath. Therefore no thermal
expansion of the capacitor itself occurs during a measurement and
the capacitance changes mainly
reflect the length changes of the sample (''small cell effect'').
Moreover, the mass whose
temperature has to be controlled is drastically reduced.
It only consists of the sample itself and a small sample holder
($\rm \simeq 50~g~Cu$). This allows for a rapid and accurate
control of the temperature using e.g. a software PID-technique.
The temperature range is restricted 
by the maximum heater current to $\rm T \lsim 200 K$.

We use platinum (Pt 103) and ''Cernox CX--1050'' temperature sensors
(Lake Shore) for temperatures above and below 100K, respectively.
The field dependence of the latter can be neglected,
since its magnetoresistance is extremely small --
the deviation is 2\% at 2.5K for H = 16 Tesla
and rapidly decreases with increasing T.
The length changes of the sample are calculated from the capacitance changes
measured by a temperature stabilized capacitance bridge
(Andeen--Hagerling) with a resolution of $\rm 5\cdot 10^{-7}\, pF$.
Thus, length changes of less than $\rm 0.01~{\AA}$
can principally be resolved. Due to mechanical vibrations etc.
the resolution is limited to $\rm \simeq 0.1 \,{\AA} $ in practice.
The thermal expansion measurements are performed in a continuous mode, i.e.
capacitance data are recorded while the
temperature is continuously increased with a small and constant rate,
usually $\rm 2-3\, mK/s$. For calibrating the dilatometer
we measured the well known~\cite{kroeger77} thermal expansion of
aluminum samples of several lengths (4 to 8 mm).
The calibration was checked
by measuring a 5mm copper sample. The relative
resolution, i.e. the scatter of the data, amounts to
$\rm \simeq 5\cdot 10^{-8}/K$ and
the deviation to the data reported in the literature~\cite{kroeger77}
is less than $\rm \simeq 1\cdot 10^{-6}/K$ for
temperatures up to 200K. In the temperature range below 45K,
which is considered in this paper only, the maximum deviation is even
less than $\rm \simeq 1\cdot 10^{-7}/K$.

For the magnetostriction measurements temperature is held constant
and length changes are detected while the magnetic field is swept from
0 up to 14 (or 16) and back to 0 Tesla with a rate of $\rm\, 2.5-4 mT/s$.
For calibration we performed magnetostriction measurements
on a 5mm silicon sample. The cell effect --
due to magnetic impurities and eddy currents
which cause stresses on the dilatometer via the
Lorentz force -- amounts up to $\rm 250 \AA$, which corresponds
to $\rm \simeq 5\cdot 10^{-6}$ for a 5mm sample.
The reproducibility of this cell effect and thus
the accuracy of the absolute values
of the magnetostriction $\rm \Delta L / L$
is better than $\rm \simeq 5\cdot 10^{-7}$.
The scatter in $\rm \Delta L / L$ amounts to
$\rm \simeq 2\cdot 10^{-9}$,
which corresponds to 0.1{\AA} as mentioned above.

\section{Results}
\label{RESULTS}

\subsection{Thermal Expansion}

Since some organic spin-Peierls compounds show a pronounced
hysteresis of the magnetic susceptibility~\cite{bloch80,bloch81}
as well as of the structure~\cite{kiryukhin95a} when changing the magnetic
field at low temperatures, all measurements of $\alpha$
were performed in the field cooled mode.
After applying the magnetic field well above $\rm T_{SP}$,
usually at $\rm T \simeq 25 K$, the
sample was cooled down to 4K (or 2K). Then
the data were taken while the temperature was
continuously increased with a rate of 2~mK/s.
In order to check the presence of hysteretic behavior we have also performed
some measurements in the zero field cooled mode, e.g.
after applying magnetic fields of H = 11T and H = 14T at 4K and
after decreasing the magnetic field from 14T to 11T at 4K.
In all cases studied the data agree with those obtained from
the field cooled measurements at the same magnetic fields.
In $\rm CuGeO_3$ the hysteresis of the lattice constants
as well as that of other
properties~\cite{poirier95a,saintpaul95a,hamamoto94a,hase93c}
in the H--T phase diagram is apparently rather weak and restricted to a very
narrow
field range, i.e. the data in Figs.~\ref{alueber} and \ref{algenau}
represent the behavior of the lattice constant $a$
in thermal equilibrium except for the
region $\rm H \simeq 12.5T$ (see below).

As shown in an earlier publication~\cite{winkelmann95} there is a very
pronounced decrease of
$\rm \alpha$ along the $a$--axis at the spin--Peierls transition
in zero magnetic field. This decrease is related
to a large negative uniaxial pressure dependence
$\rm dT_{SP}/dp_a \simeq -4$K/GPa on the one hand and to a spontaneous
lengthening of the $a$--axis in the dimerized
phase on the other hand. Fig.~\ref{alueber} gives an
overview of the changes of the thermal
expansion coefficient along the $a$--axis occurring as a function
of magnetic field. A pronounced field dependence of
$\rm \alpha$ is present only below 20K, i.e.
in the spin--Peierls phase.
First of all the transition temperature $\rm T_{SP}$
reduces with increasing field, a result
known e.g. from measurements of the
susceptibility~\cite{hamamoto94a,kremer95}.
In addition, a change of the
anomaly of $\rm \alpha$ as a function of the magnetic field is
apparent from Fig.~\ref{alueber}. In all cases
the phase transition shows up by a pronounced decrease of
$\rm \alpha$. However, note that in Fig.~\ref{alueber}
the anomaly for H = 14T is distinctively smaller than that for H = 8T,
whereas the sizes of the anomalies for H = 8T and H = 0T are very similar.

This nonlinear change of the size of the anomalies is related to the
different low temperature phases occurring as a function of magnetic field.
In H = 8T the anomaly still arises from the spin--Peierls transition
between the U and D
phases~\cite{poirier95a,hamamoto94a,kiryukhin95b},
whereas the anomaly of $\rm \alpha$ for H = 14T is due to a
phase transition between the U the I phases.
Although the anomaly for H = 14T is reduced in size,
a clear structural change between the U and I phases
of $\rm CuGeO_3$ is inferred from these data.
Moreover,  this transition is strongly pressure dependent
and leads to a spontaneous lengthening of the $a$--axis,
similar to that in zero magnetic field.

Fig.~\ref{algenau} shows an expanded view of the
temperature and magnetic field dependence of $\rm \alpha$
up to 17.5K and for $\rm 0T \le H \le 16$T.
With increasing magnetic field the anomaly of
$\rm \alpha$ is systematically shifted to lower temperatures, reflecting the
decrease of the transition temperature with increasing magnetic field.
This decrease amounts to about
$\rm T_{SP}(0) - T_{SP}(H) \simeq 2 K$ for
fields up to H = 11T and to $\rm \simeq 4 K$
for H = 16T, respectively. Considering the shape and size
of the anomalies the curves can be clearly
classified into three groups: (i) Up to 12 Tesla the
anomalies remain nearly unchanged.
(ii) A very pronounced decrease of ''$\Delta\alpha$'' as a function of
the magnetic field is present in the rather small field range
between 12T and 13T. Besides the strong field dependence this second
group of curves is characterized by additional anomalies occurring below
10K. As shown in Fig.\ref{al1213} very strange thermal expansions
indicating the presence of several phase transitions as a function of
temperature are found in the entire field range between 12T and 13T,
most pronounced for H = 12.5T (see Fig.~\ref{al1213}).
(iii) In the field range between 13T and the maximum field of 16T
again only one transition is present. The sizes of the anomalies
are much smaller than those at low fields
and they further reduce with increasing magnetic field.
Despite of the limited field range for the investigation of this third
group a stronger magnetic field dependence of the anomaly size
than in the first group ($\rm H < 12T$) is apparent even from the raw data.

The different groups of curves reflect the different kinds of phase
transitions. Below 12 Tesla there is a transition between the U and D
and above 13 Tesla between the U and I phases.
In the field region $\rm 12T \le H \le 13T$ the three phase boundaries
U/D, U/I, and D/I meet in a tricritical point.
At temperatures below this tricritical point
two transitions are expected with increasing magnetic field (D/I and I/U).
A sequence of transitions is also possible and present
(see Fig.~\ref{al1213}) as a function of temperature pending on the
details of the D/I phase boundary, which is nearly
horizontal in the H--T phase diagram, i.e. occurring at a nearly
constant magnetic field.

From the data presented so far it becomes apparent that
all transitions between the three phases of $\rm CuGeO_3$
lead to pronounced anomalies of the thermal expansion coefficient,
i.e. each phase transition causes spontaneous strains.
It should be mentioned that a
dimerization alone, which is characterized by the
development of alternating distances between nearest neighbors,
i.e. an antiferro distortion, does not necessarily lead
to spontaneous strains. Especially, within the CF theory
pressure dependencies of $\rm T_{SP}$
and thus anomalies of the thermal expansion coefficients
are obtained only, if one adds an anharmonic coupling between
elastic degrees of freedom and zone boundary
phonons~\cite{bray80,lepine86}.
Such a coupling is of course no general property of the
spin--Peierls transition and thus the pressure dependencies strongly
differ for the spin--Peierls compounds~\cite{bray80,lepine86}.

\subsection{Magnetostriction}
\label{MAGNETOSTRIC}

Whereas the measurements of thermal expansion are a very sensitive probe
of the U/D and U/I phase transitions, it is rather difficult to
determine the D/I phase boundary since the temperature dependence
of the latter is very small.
Therefore we also performed measurements of the magnetostriction,
i.e. measurements of the magnetic field induced
changes of the lattice constant $a$ at a fixed temperature, where the
D/I-phase boundary is crossed almost perpendicular.

In Fig.~\ref{ms} we show $\Delta a{\rm (H)/}a({\rm H=0)}$ obtained
with increasing magnetic field at several temperatures between
3 and 18K.
At the lowest temperature there is only an extremely
small magnetostriction in the D phase and at
$\rm H_{D/I} \simeq 12.5\, Tesla$ a large jumplike decrease
of the lattice constant $a$ occurs reflecting the
first order phase transition between the D and the I phase.
With increasing temperature the ''jump'' of the
lattice constant at the D/I transition strongly decreases indicating that
the phase transition
gradually changes from a discontinuous to a continuous one.
Note, however, that $\rm H_{D/I}$ and
the total magnetostriction up to 14 Tesla, i.e. $a{\rm (14T)-}a{\rm (0T)}$,
remain roughly constant below 11K.
Moreover, a large magnetostriction is found in the D phase in this
temperature range.
A comparison with the thermal expansion data in Fig.~\ref{algenau} shows
that all magnetostriction curves obtained in this temperature range
reflect differences of the lattice constant between
the D and the I phases. The gradual change from a
clearly first order transition to a continuous one
can also be extracted from the hysteresis of the magnetostriction around
$\rm H_{D/I}$ (see inset of Fig.~\ref{ms}).
At low temperatures a small hysteresis with a maximum value of about
0.2 Tesla (at  T$\rm \simeq$ 3K) is determined when comparing the measurements
with increasing and decreasing magnetic field.
With increasing temperature the amount of this irreversibility
systematically decreases and vanishes for $\rm T \gsim 11K$.

At temperatures between $\rm \simeq 11K$ and 14.5K,
i.e. for $\rm T_{SP}(0T) > T > T_{SP}(12T)$
(see Fig.~\ref{algenau}) the field driven transitions are no
longer between the D and I phases.
The still very pronounced and continuous decrease of the lattice constant $a$
up to a critical field now reflects
the continuous phase transition between the low field D
and the high temperature U phase of $\rm CuGeO_3$.
In contrast to the behavior at lower temperatures
both the transition field and the total
magnetostriction rapidly decrease with increasing temperature.
The first observation arises from the decrease of $\rm T_{SP}$ as a function
of the magnetic field (see Fig.~\ref{algenau}), which
implies a decrease of $\rm H_{D/U}$ with increasing
temperature. The second observation reflects the temperature dependence of the
spontaneous strain, i.e. the continuous increase of the structural
difference between the D and U phases with decreasing
temperature.

At temperatures above 14.5K no phase transition is found.
However, a finite magnetostriction is clearly observable also in the
U phase. Note that in contrast to the findings at lower temperatures the
lattice constant $a$ now {\em increases} with increasing magnetic field.
This magnetostriction in the U phase
is not related to fluctuations of the spin--Peierls
order parameter. It measures the magnetoelastic coupling in
$\rm CuGeO_3$, which is a precondition for the occurrence  of a spin--Peierls
transition~\cite{buechner96a}. It should be mentioned that this
positive magnetostriction is found in the entire temperature range studied,
i.e. up to 80K. A detailed discussion of the magnetoelastic coupling in the
uniform phase as extracted from the magnetostriction
and its anisotropy will be given in a forthcoming publication.
The finite and slightly
temperature dependent magnetostriction in the U
phase implies a field dependence of the thermal expansion for
$\rm T > T_{SP}$. However, in agreement with the findings presented in
Fig.~\ref{alueber} this field dependence is estimated
to be extremely weak. As an upper limit
for the difference $\rm \alpha(0T) - \alpha(14T)$
we obtain $\rm 2\cdot10^{-8} /K$ for $\rm 20K < T < 80K$
corresponding to a relative change of $\rm \alpha$ smaller than one percent.
However, this finite magnetostriction in the
U phase has to be taken into account for a quantitative comparison of
the results of thermal expansion and magnetostriction measurements at low
temperatures (see below).

\section{H--T phase diagram}
\label{PHADI}

From the measurements of magnetostriction and thermal expansion
it is possible to determine a complete and precise
H--T phase diagram of $\rm CuGeO_3$ since both quantities
show pronounced anomalies at the different phase transitions.
The anomalies
of the magnetostriction give the fields at the D/I and D/U
phase transitions at different temperatures
while those of $\alpha$ give the temperatures at
the U/D and U/I transitions at different magnetic fields.
It is apparent from the raw data presented so far that
(i) the phase boundaries in $\rm CuGeO_3$ follow
the characteristic course expected
for the spin-Peierls transition and that (ii) our observations are in rough
agreement with the findings from other properties, e.g.
the magnetization~\cite{hase93c,hamamoto94a}. We did, however, not observe
any anomaly of the thermal expansion or magnetostriction
at H = 8 Tesla, where Poirier et al. found evidence for an
additional phase boundary from ultrasonic studies~\cite{poirier95a}.

The $\lambda$--like shape of the anomalies
of $\rm \alpha$ but also of the specific heat~\cite{liu95b}
signal the presence of pronounced fluctuations in $\rm CuGeO_3$,
which have to be taken into account
when determining the transition temperatures.
In literature quite different ways have been used
to define $\rm T_{SP}$ from the anomaly of the
specific heat~\cite{oseroff95,liu95b}.
Ignoring the fluctuations and describing the
anomaly as a broadened mean field step as in Ref.~\cite{oseroff95}
one would obtain a transition temperature of $\rm \simeq 14.5K$ for
our crystal. However, this description is apparently not appropriate
for our more homogeneous sample.
If one assumes a (nearly) symmetric shape of the anomaly on a reduced
temperature scale and a linear regular part~\cite{liu95b},
i.e. the behavior expected for critical fluctuations,
the transition temperature is distinctively smaller. It is
found close to the maximum of $\rm C_p$ or the minimum of $\rm \alpha$,
i.e. at 14.16K for our crystal. However, the experimentally observed
anomalies of the thermodynamic properties of $\rm CuGeO_3$
all show a pronounced asymmetry even close to ''$\rm T_{SP}$''.
Therefore the temperature at the minimum of
$\rm \alpha$  only yields a lower limit for the
transition temperature.
In this paper we have defined $\rm T_{SP}$ at the maximum slope
of the anomaly yielding $\rm T_{SP} = 14.35$,
which lies in between the two extreme values mentioned above.
We emphasize that, despite of this rather large uncertainty of the
absolute value, the decrease of $\rm T_{SP}$ as a function of H
is obtainable with very high precision
(better than $\rm \pm 0.05 K$) from our data, e.g. from the shift of
the minimum in $\rm \alpha$, since the shape of the anomalies does
not change significantly (see below).

In Fig.~\ref{btphadi} the H--T phase diagram
of $\rm CuGeO_3$ for fields parallel to the $a$--axis
is displayed. The transition fields
measured via the magnetostriction
are defined in a similar way as $\rm T_{SP}$, i.e.
at the maximum value of the field derivative.
As visible in Fig.~\ref{btphadi}
the boundaries obtained from the two experimental methods
perfectly agree. For the transition fields between
the D and I phases two values are given representing
the hysteresis at this phase transition.
The  lower (higher) value corresponds to the transition field found with
decreasing (increasing) magnetic field. The inset of
Fig.~\ref{btphadi} shows the D/I phase boundary on a smaller field scale.
The hysteresis can be clearly resolved at least up to about 10K.
This hysteresis and the shape of the
anomalies of the magnetostriction (see Fig.~\ref{ms}) show that the field
driven D/I transitions are of first order. The shapes of the
additional anomalies in the thermal expansion (Fig.~\ref{al1213}), which show
up very close to this phase boundary, also indicate a weakly first order D/I
transition as a function of temperature.

Similar to the experimental results found in the organic spin--Peierls
compounds~\cite{bonner87} the
H--T phase diagram of $\rm CuGeO_3$ does not support
the soliton theories of the spin--Peierls transition.
Most of the arguments given in Ref.~\cite{bonner87}, which favor the
traditional theories, e.g. the CF theory, in the case of TTF--BDT(Au)
also hold for $\rm CuGeO_3$.
For example the soliton theory predicts a continuous D/I transition, whereas
experimentally a first order transition is found.
Moreover, the D/I transition is predicted to occur at
$\rm g\mu_B H_{D/I} \simeq 0.6 E_0$, where
$\rm E_0$ denotes
the spin excitation gap at zero temperature~\cite{nakano80,nakano81}. Using
the excitation gap derived from inelastic neutron scattering
studies~\cite{fujita95a}, the D/I transition should occur at
$\rm H_{D/I}\simeq 8.5 T$,
which is significantly smaller than the experimental value.

The experimental data agree much better with the calculations of
Cross~\cite{cross79b} and the earlier treatment of Bulaevskii et
al.~\cite{bulaevskii78}. A simple comparison to these calculations
is possible by
considering firstly the decrease of $\rm T_{SP}$ at small magnetic fields
and secondly
the tricritical point, where the U/D, U/I, and I/D phase boundaries meet.
Both theories yield in leading order  a quadratic decrease of
$\rm T_{SP}(H)$ with H,
i.e. $\rm (T_{SP}(H) - T_{SP}(0)) / T_{SP}(0) = - \tau
\cdot \left( \frac{g\,H}{2\,T_{SP}(0)}\right)^2$.
The predicted quadratic field dependence is confirmed by our data
up to about 6T. The prefactor amounts to $\rm \tau = 0.21 K/T$,
which agrees with the data reported by Poirier et al.~\cite{poirier95a}.
Taking into account the g--factor of 2.15 for fields parallel to the
$a$--axis~\cite{ohta94} the theoretical treatments of Bulaevskii et al.
and Cross yield $\rm \tau = 0.19 K/T$ and $\rm \tau = 0.16 K/T$,
respectively.
The tricritical point is derived from our measurements at
$\rm H_c = 12.25\,Tesla$ and $\rm T_c = 11K$ (see Fig.~\ref{btphadi}).
The reduced values  $\rm \frac{T_c}{T_{SP}(0)} = 0.77$ and
$\rm \frac{g\,H_c}{2\, T_{SP(0)}} = 0.92 T/K$
can directly be compared to theory.
Bulaevskii et al. find ratios of 0.63 and 1.08~\cite{bulaevskii78},
whereas Cross calculates $\rm 0.77$ and $\rm 1.03$~\cite{cross79b} for
the reduced tricritical temperature and field, respectively.
Apparently, our data very well agree with the prediction of Cross, especially
with regard to $\rm T_c$. The deviations from the calculations
of Bulaevskii are somewhat larger.
This is expected for quite general reasons,
since more reliable results are obtained from Cross'
Luther Peschel treatment of the one--dimensional spin excitations
than from the Hartree Fock approach used by Bulaevskii et al..

In Fig.~\ref{btphadi} the phase boundaries
calculated by Cross are compared to our
experimental findings. As mentioned above there are
clear discrepancies. However, the shape of the boundaries is very
similar and there is an easy way to obtain a perfect agreement between
the theoretical curve and the experimental U/D phase boundary.
As also shown in Fig.\ref{btphadi} the experimental data
follow a theoretical curve, when the reduced field scale
is divided by 1.12 (dotted line). The ''effective'' magnetic field,
which acts in $\rm CuGeO_3$, seems to be larger than
that calculated. Cross himself already gives
some possible reasons for this kind of deviation~\cite{cross79b}. For
example, using a too small zero--temperature susceptibility
to derive the phase diagram causes an ''effective'' magnetic
field that is also too small. Thus, the discrepancy
between theoretical and experimental H--T phase diagrams
might be related to the striking discrepancy
between the measured and calculated
susceptibility of $\rm CuGeO_3$ in the U phase.
The latter has been attributed to a next nearest
neighbor exchange~\cite{castilla95,riera95} in the quasi--one dimensional
chains in $\rm CuGeO_3$. Recently, first calculations of the H--T phase diagram
within a model incorporating a next nearest neighbor exchange have been
published~\cite{riera96a}. At the present stage these calculation
do, however, not improve the agreement with the experimental data.

Since particular properties of $\rm CuGeO_3$ might
influence the H--T phase diagram, a comparison
to the findings in other spin--Peierls compounds seems
worthwhile~\cite{northby82b}. As shown by Hase et al.~\cite{hase93c,hase94}
the H--T phase diagrams of all spin--Peierls compounds
roughly coincide on reduced temperature and field scales.
However, some small but systematic differences are present.
This is most clearly visible for
the D/I phase boundary, which is located
at larger (reduced) magnetic fields
in the organic compounds than in $\rm CuGeO_3$.
Despite of the rather large scatter of the data for the organic compounds
the same trend is also found when comparing the D/U boundaries of
the organic compounds to our data of $\rm CuGeO_3$.
Thus, an enhanced ''effective'' magnetic field,
which destabilizes the D phase of $\rm CuGeO_3$, can also
be inferred from a comparison to the
organic spin--Peierls compounds.

In order to compare the phase boundaries to the I phase
one has to take into account
a particular property of the organic spin--Peierls compounds.
A pre--existing soft phonon, which
stabilizes the commensurate lattice deformation~\cite{bulaevskii78,cross79b},
seems to be a characteristic feature of all these systems.
In $\rm CuGeO_3$ the experimentally observed
field at the D/I transition is rather low compared to the
calculated one. Moreover, the transition temperatures at the
U/I transitions are larger and less field dependent than
predicted by theory.  Thus, the H--T phase diagram yields
no evidence that details of the phonon spectrum favor the
D phase compared to the other phases.
In agreement with this conclusion drawn from the H--T phase diagram
no pre--existing soft phonon has been
observed so far in $\rm CuGeO_3$.

\section{Field dependence of the thermal expansion and the spin--Peierls
order parameter}
\label{FIELDDEP}

The presented data of the magnetostriction
and the thermal expansion in magnetic fields do not only allow
for an accurate determination of the phase diagram.
Information on thermodynamic properties
of the phase transition as well as information about the
field and temperature dependence of the structure are also obtained
from these high resolution measurements of the lattice constant $a$.
Before we are going to discuss the field dependencies in detail we
shortly remind
on the basic conclusions drawn from the zero field data in the
following section. In addition, all properties,
whose field dependencies will be analyzed in the subsequent
four chapters, are defined in this section.

\subsection{Zero magnetic field}
\label{ZERO}

The main conclusions from the anomalies of the thermal expansion coefficient
at $\rm T_{SP}$ in zero magnetic field have already been presented
in a previous publication~\cite{winkelmann95}. From general thermodynamic
arguments an anomaly of $\rm \alpha$ is expected
for each phase transition with a finite pressure dependence
of the transition temperature.
This yields not only for mean field jumps, where
the anomalies of $\rm \alpha$ and the specific heat $\rm C_p$ are related via
the Ehrenfest relation, but also for phase transitions strongly affected
by fluctuations~\cite{pippard64} as the spin--Peierls transition in
$\rm CuGeO_3$~\cite{winkelmann95,liu95b}. Assuming a finite pressure
dependence of $\rm T_{SP}$ one expects the same temperature dependence
of the leading singular parts of $\rm C_p$ and $\rm \alpha$
at the phase transition. Quantitatively both quantities
are related by a scaling factor measuring the (positive
or negative) uniaxial pressure dependence of the transition temperature.
Thus, as long as one treats the anomalies of the thermal expansion coefficient
($\Delta \alpha$) and the specific heat
($\Delta C$) in a similar way, one can determine the uniaxial
pressure dependence of the transition temperature
in the limit of vanishing pressure.
The relationship reads
\begin{equation}
\rm
\label{Ehri}
\left.\frac{\partial T_{SP}}{\partial p}\right|_{p\rightarrow 0} =
V_m  T_{SP}\cdot\frac{\Delta\alpha}{\Delta C} \, ,
\end{equation}
where $\rm V_m$ denotes the volume per mol.
Experimentally it is found that in $\rm CuGeO_3$ the similarity between
$\rm C_p$ and $\alpha$ is restricted to a very narrow temperature range
$\rm |T_{SP}-T| \lsim 0.2K$,
where sample inhomogeneities may strongly alter the
temperature dependence. Therefore it is not possible
to extract the critical behavior of the
specific heat in $\rm CuGeO_3$~\cite{liu95b} from the thermal expansion
data. Nevertheless, Eq.~\ref{Ehri} is valid for $\rm CuGeO_3$, since
the hydrostatic pressure dependence calculated
from the thermal expansion and specific heat~\cite{winkelmann95} perfectly
agrees with the hydrostatic pressure dependence that is directly determined via
susceptibility measurements at finite pressures~\cite{takahashi95}.

To obtain a quantitative measure of the anomaly
size we consider the largest deviation of
the measured thermal expansion from the extrapolated
behavior above $\rm T_{SP}$ ($\rm \alpha_{extr.}$, solid line
in Fig.~\ref{alueber}) $\rm \Delta \alpha
\equiv max(\alpha(T<T_{SP})-\alpha_{extr.}$), i.e. the height
of the asymmetric peaklike anomaly. Note that $\Delta\alpha$ does
not correspond to a real jump of $\rm \alpha$ in the sense of a mean field
theory. Below $\rm T_{SP}$ the anomalous contribution to $\rm \alpha$,
i.e. the difference between the measured and extrapolated
temperature dependence
$\delta\alpha(T) \equiv \alpha(T) - \alpha_{extr.}(T)$,
strongly differs from that of the specific heat. This is
found experimentally, but also expected from e.g. CF theory.
$\rm C(T<T_{SP})$ is
dominated by the magnetic excitations and therefore shows
activated behavior at low
temperatures~\cite{sahling94,liu95b,oseroff95}, whereas $\rm \delta\alpha$
is closely related to the structural
order parameter ($\rm Q$), i.e. the dimerization of the lattice.
By definition it measures the development of a spontaneous
strain ($\rm \epsilon$) in the D phase~\cite{winkelmann95,lorenzo94,harris94}.

To describe the temperature dependence of $\epsilon$ and $\delta\alpha$
and their relationship to $\rm Q$ one can start
from the usual Ginzburg--Landau expansion of the free
energy. Taking the expression given by Cross and Fisher~\cite{cross79a},
adding the elastic energy $\rm 1/2c\epsilon^2$ ($\rm c=elastic$ constant)
and assuming a coupling between dimerization and
lattice strains $\rm \mu Q^2\epsilon$ ($\mu$ = const.),
which is quite common for structural phase
transitions~\cite{brucecowley}, the free energy reads:
\begin{equation}
\rm
\label{landi}
F = \gamma \left[ 1 - \frac{T}{T_{SP}} -
\frac{1}{2} \eta \left(\frac{H}{T_{SP}}\right)^2\right]Q^2
+ \frac{1}{2} \nu Q^4 + \mu Q^2\epsilon + \frac{1}{2}c\epsilon^2
\end{equation}
The constants $\gamma$, $\eta$, and $\nu$
are related to the mean field thermodynamic properties below $\rm T_{SP}$
and the leading order of the decrease of $\rm T_{SP}$ in magnetic
fields~\cite{cross79a}.
In thermodynamic equilibrium without external
stress, i.e. for $\rm {\partial F \over \partial \epsilon}=0$,
it is easy to obtain the relationship between $\epsilon$
and the order parameter from Eq.~\ref{landi}.
The spontaneous strain is proportional to the
square of the order parameter~\cite{winkelmann95,harris94}
and thus the anomalous contribution of
the thermal expansion represents the temperature derivative of $\rm Q^2$:
\begin{equation}
\rm
\label{scali}
Q^2(T) = -\frac{c}{\mu}\epsilon(T) \equiv -\frac{c}{\mu} \cdot
\int\limits_{T_{SP}}^T\delta\alpha(T') dT'
\end{equation}

In Fig.~\ref{epsop} we show that in zero field the scaling between
the spontaneous strain obtained from the thermal expansion coefficient
and the intensity of a superstructure reflection from neutron scattering
data~\cite{regnault96a} is fulfilled
in the entire temperature range below $\rm T_{SP}$ (see also
Ref.~\cite{harris94}). The intensity of the superstructure reflection is
proportional to the square of the dimerization, i.e.
of the structural order parameter. Moreover, it is apparent
from Fig.~\ref{epsop} that the thermal expansion yields a
high resolution measurement of $\rm Q^2(T)$,
since the scatter of these data
is much smaller than that of the diffraction experiments.

At this point we mention that it is also possible to determine
the temperature dependence of the spin--Peierls order parameter
in external fields from Equ.~\ref{scali}.
In this case one has to define the anomalous contribution
of $\alpha$ with respect to the extrapolated behaviour of
$\alpha$ measured in the corresponding field.
For quite general reasons a finite field dependence of
$\alpha$ in the U phase, i.e. a finite magnetostriction,
is expected due to the magnetoelastic coupling
which is a precondition for the
occurrence of a spin--Peierls transition.
This magnetoelastic coupling
also implies a magnetic field dependence
of $\rm \alpha_{extr.}$. However, in our analysis
of $\rm Q(T,H)$ which we will present in the
subsequent sections we will neglect this
field dependence of $\rm \alpha_{extr.}$ for two reasons.

First, as shown in Fig.~\ref{alueber} the field dependence
of $\alpha$ in the U phase is extremely small (not resolvable).
Note that this does not contradict a finite magnetostriction,
since the magnetic part of $\alpha$ does not scale
with the magnetostriction itself but with its temperature
dependence, which is very weak in the U phase as mentioned in
section~\ref{MAGNETOSTRIC}.
Second, although $\rm \alpha_{extr.}$ is not known exactly,
it is apparent from Fig.~\ref{alueber} that it is
much smaller than the anomalous contribution $\delta\alpha$.
Thus, any reasonable choice of the background as a function
of temperature and field yields the same anomalous contribution
$\delta\alpha$ within our resolution. Therefore in the following we will
assume a field independent $\rm \alpha_{extr.}$.
Up to now there are only few diffraction data of $\rm CuGeO_3$ in
magnetic fields available to check our assumption.
The Inset of Fig.~\ref{epsop} compares the spontaneous strain
in a field of 10 Tesla to the intensity of a superstructure
reflection in a field of 9.85 Tesla~\cite{regnault96a}.
Although there are small differences in the fields as well as
their orientations with respect to the crystal axes
the agreement of the data is quite good.
We mention that the same scaling factor was used for comparing the
zero field and field data, respectively.

Let us now turn to the discussion of the temperature dependence
of the spin--Peierls order parameter in zero field.
In Fig.~\ref{bcsop} we compare the experimentally determined
$\rm Q^2(T)/Q^2(0K)$ to a theoretical BCS temperature
dependence (dotted line),
which has been found for other spin--Peierls compounds~\cite{moncton77}.
Obviously, there are strong deviations in the entire temperature
range below $\rm T_{SP}$  for $\rm CuGeO_3$.
For $\rm T \rightarrow T_{SP}$ this is of course due
to the fluctuations of the order parameter.
As a further consequence of these fluctuations
the transition temperature is reduced compared to a
hypothetical mean field transition
temperature ($\rm T_{SP}^{mf}$).
Since the mean field temperature dependence
of Q depends on $\rm T_{SP}$
the fluctuations may also explain the deviations between the
measured $\rm Q^2(T)$ and the dotted line in Fig.~\ref{bcsop} at low
temperatures. Vice versa, assuming a BCS mean field temperature
dependence at low temperatures one can estimate the
decrease of the transition temperature due to fluctuations.
A corresponding analysis is shown in Fig.~\ref{bcsop}. Below
about 10K the experimentally found $\rm Q^2(T)$ follows a BCS temperature
dependence with a $\rm T_{SP}^{mf} \simeq 15.8K$,
i.e. $\rm (T_{SP}^{mf}-T_{SP})/T_{SP}^{mf} \simeq 10\%$.
Although there is a large uncertainty and in addition to that
no real physical meaning of $\rm T_{SP}^{mf}$, we
emphasize that a quantitative comparison
to mean field theories
has in principle to consider a decrease of $\rm T_{SP}$ which is
due to fluctuations.

Close but not too close to the phase transition the temperature dependence
of the spontaneous strain is well described by a
power law $\rm \epsilon \propto (T_{SP} - T)^{2\beta_\epsilon}$
(Fig.~\ref{bcsop}, see also Ref.~\cite{winkelmann95,harris95}). From a fit to
our data we find $\rm 2\beta_{\epsilon} = 0.61(5)$.
The large error in the exponent is a direct
consequence of the uncertainty in the absolute value of $\rm T_{SP}$ discussed
above. Renormalization group theory predicts an exponent $\beta=0.325$ for the
order parameter of the three dimensional structural transition
in $\rm CuGeO_3$ (universality class 3d Ising). This
prediction is confirmed by our data when assuming the
scaling between $\epsilon$ and $\rm Q^2$ (see Eq.~\ref{scali}).
However, note that the large error bar of $\rm 2\beta_{\epsilon}$
prevents a further discrimination between different three dimensional
universality classes (XY, Heisenberg). Moreover,
deviations from the power law behavior of the spontaneous strain occur
very close to $\rm T_{SP}$.
This is only partially due to sample inhomogeneities, since such
deviations are also expected for a more general reason.
As mentioned above the critical behavior of the thermal expansion coefficient
for temperatures very close to $\rm T_{SP}$
is expected to scale with that of the specific heat.
Therefore the critical behavior of $\rm \epsilon = \int\delta\alpha$
should change very close to $\rm T_{SP}$ and
finite values of $\rm \int\delta\alpha$ above $\rm T_{SP}$ are expected.

In the following four sections we will investigate the field
dependence of the quantities defined above, i.e.
the field dependence of:\\
(i) the anomaly size $\Delta\alpha$,\\
(ii) the critical exponent $2\beta_{\epsilon}$,\\
(iii) the spontaneous strain $\epsilon$, and\\
(iv) the temperature dependence of the order parameter $\rm Q$.

\subsection{Field dependence of $\rm \Delta\alpha$}
\label{DELTAALPHA}

As can already be inferred from Figs.~\ref{alueber} and \ref{algenau},
the size of the anomalies of $\rm \alpha$ markedly decreases
as a function of H. In Fig.~\ref{peaks} $\rm \Delta\alpha$ is plotted
as a function of the magnetic field. The correlation
between this field dependence of $\rm \Delta\alpha$ and the
phase diagram is apparent. A dramatic change of $\rm \Delta\alpha$
occurs exactly at the field, where the
character of the phase transition changes. Therefore a clear
discrimination between U/D and U/I transitions is possible, when
considering the
temperature dependence of the lattice constants.
Furthermore, our data indicate significant structural differences
between the D and I phases. The much smaller anomalies of $\rm \alpha$
at the U/I transitions mean that the
spontaneous strain in the I phase is strongly reduced
compared to that in the D phase (see below).
Besides this pronounced change of $\rm \Delta\alpha(H)$ at the tricritical
point the data in Fig.~\ref{peaks} also show a smaller decrease
of $\rm |\Delta\alpha|$ with increasing field for both U/D and
U/I transitions. Although we could follow the latter
only within a restricted field range, our data suggest a
rather strong field
dependence of $\rm \Delta\alpha$ at the U/I phase boundary,
which amounts to about $\rm 5\% /Tesla$.
This rather strong decrease, which occurs
at a roughly constant $\rm T_{SP}$, indicates the presence of
continuous structural changes within the I phase as a function of H.

At the U/D phase boundary we find a much weaker
field dependence of $\rm \Delta\alpha$,
though the transition temperature
decreases stronger. Up to $\rm H=12.15~Tesla$
the reduction of $\rm \Delta\alpha$ amounts only to about 10\%.
Note that the decrease of $\rm |\Delta\alpha|$ obtained in the entire field
range of U/D transitions is much smaller than the decrease
of the specific heat anomaly found for a rather small field of
6 Tesla~\cite{liu95b}. Moreover, as
visible in Fig.~\ref{peaks} $\rm |\Delta \alpha|$ shows a
nonlinear decrease with increasing field similar to that of the
transition temperature $\rm T_{SP}(H)$. Indeed the decreases of both
quantities can empirically be related to each other.
The ratio $\rm \Delta\alpha / \sqrt{T_{SP}(H)}$
(right part of Fig.~\ref{peaks}) is a constant value
for all magnetic fields,
i.e. $\rm \Delta\alpha(H) \propto \sqrt{T_{SP}(H)}$.

The size of the anomaly of $\rm \alpha$ is related to the
(uniaxial) pressure dependence of $\rm T_{SP}$ (Eq.~\ref{Ehri}.),
which strongly differs for different spin--Peierls
compounds~\cite{reviewbray}.
Thus, $\rm \Delta\alpha$ itself is not predicted by theory.
However, the field dependence of $\rm \Delta\alpha$ can be compared
-- at least qualitatively -- to theoretical predictions.
The universal H--T phase diagram of spin--Peierls compounds
implies a similar universality for a single compound
when studied at finite pressure. This is valid at least in the limit of
vanishing pressure, which is sufficient to discuss our data.
A pressure induced change of $\rm T_{SP}(H=0T)$
also affects the scaling of the field axis in the phase diagram.
To take this into account $\rm \partial T_{SP}/\partial p$ has to change
as a function of H:
\begin{eqnarray}
\label{Druecki}
\rm
\left.\frac{\partial T_{SP}(H,p)}{\partial p}\right|_{p \rightarrow 0} &=&
\left.\frac{\partial T_{SP}(0,p)}{\partial p}\right|_{p \rightarrow 0} \cdot
\left( \frac{T_{SP}(H)}{T_{SP}(0)} - \frac{H}{T_{SP}(0)}\cdot
\frac{\partial (T_{SP}(H))}{\partial H} \right)\\
\label{Druecki1}
&\simeq&
\left.\frac{\partial T_{SP}(0,p)}{\partial p}\right|_{p \rightarrow 0} \cdot
\left( 1 + \tau \frac{g^2 H^2}{4 T^2_{SP}(0)}\right) {\rm (for~~H \leq 6T)}
\end{eqnarray}
Note that the expression in the brackets of Eq.~\ref{Druecki1},
which is derived from the
quadratic decrease of $\rm T_{SP}(H)$ (see section~\ref{PHADI}),
is larger than 1. This means that the absolute value of the
pressure dependence of $\rm T_{SP}$
increases with increasing magnetic field.
Combining Eq.~\ref{Ehri} and~\ref{Druecki}
the field dependence of $\Delta\alpha$ is given by:
\begin{equation}
\label{alphi}
\frac{\Delta\alpha(H)}{\Delta\alpha(0)} =
\frac{\Delta C(H)}{\Delta C(0)}
\cdot
\left(1 - \frac{H}{T_{SP}(H)}\cdot
\frac{\partial (T_{SP}(H))}{\partial H} \right)
\end{equation}
It is apparent from this equation that one expects a rather small
change of $\rm \Delta \alpha$ as a function of the magnetic field.
Especially, the decrease of the relative anomaly size
$\rm \Delta \alpha (H) / \Delta\alpha(0)$
is expected to be much weaker than that of the specific heat.
However, a detailed knowledge
of the field dependence of the specific heat is necessary
to analyze the simple scaling we find between
$\rm \sqrt{T_{SP}}$ and $\rm \Delta\alpha$ (Fig.~\ref{peaks}).
Simple models for $\rm \Delta C(H)$, e.g.
based on the free energy given in~\cite{cross79a}, are not
sufficient. Assuming a mean field behavior
$\rm \Delta C(H) \propto T_{SP}(H)$ one even obtains an
increase of $|\Delta\alpha|$ with increasing field,
which is in disagreement with the data.
Studies of the specific heat in high magnetic fields are in progress
and a detailed comparison of $\rm \Delta C(H)$ and $\rm \Delta \alpha(H)$
obtained at the same crystal will be given in a forthcoming publication.
At present we conclude that the rather small field dependence of
$\rm \Delta \alpha$ for fields below 12 Tesla shown in Fig.~\ref{peaks}
is in qualitative agreement with the field dependence estimated
from thermodynamic relations.

\subsection{Critical behavior of
the spontaneous strain in external fields}

In zero magnetic field the spontaneous strain $\epsilon$
close to $\rm T_{SP}$ follows a power law expected
for the squared order parameter (see Fig.~\ref{bcsop}).
In Fig.~\ref{critis} we show $\rm \epsilon $ as a function of the reduced
temperature $\rm 1 - T/T_{SP}(H)$ for several magnetic fields
in a double logarithmic scale. From these plots a power
law behavior in a rather field independent temperature range
$\rm 0.7< T/T_{SP}<0.95$ is found for all fields studied except those
very close to the tricritical point, where two transitions occur
(see Fig.~\ref{al1213}). Moreover, it is apparent
from Fig.~\ref{critis} that for low magnetic fields the slope in the double
logarithmic scale is roughly constant,
i.e. the exponent $\rm 2\beta_\epsilon$ is constant.
There is, however, a pronounced difference between the
low field data and those at 15 and 16 Tesla, where a smaller exponent
is present.

The field dependence of $\rm 2\beta_\epsilon$
is plotted in the lower part of Fig.~\ref{critis}.
As mentioned above and displayed in the figure
there is a rather large systematic error ($\rm \pm 0.05$)
of the absolute values due to
the uncertainty in the absolute value of $\rm T_{SP}$ at a given field.
This uncertainty does, however, not affect
the field dependence of $\rm 2\beta_\epsilon$, since
the field induced decrease of $\rm T_{SP}$
is extractable with much higher accuracy from our data.
In other words, an alternative definition of $\rm T_{SP}$ in zero
magnetic field causes a shift of $\rm 2\beta_{\epsilon}$
to higher or lower values,
but the field dependence of $\rm 2\beta_\epsilon$
shown in the lower part of Fig.~\ref{critis}
remains nearly unchanged.
For the U/D transitions at low fields a large exponent
of about 0.6 is found, which decreases only slightly
with increasing field. Within the  error this value
is consistent with that expected for
a 3d Ising transition ($\rm \beta = 0.325$) as
discussed for the zero field data.

The critical exponents of the U/I transitions are significantly
smaller than those of the U/D transitions and show a rather strong
increase with increasing magnetic field.
The interpretation of the findings for the U/I transition
is more complicated for two reasons.
First the nature of the I phase is still
discussed controversially~\cite{kiryukhin95b}
and there are to our knowledge no theoretical predictions for
the critical exponents. Second our studies are restricted to a
field range rather close to the tricritical point
where theory predicts a smaller exponent $\rm \beta = 0.25$.
This proximity might be the reason for the systematic increase
of $\rm 2\beta_{\epsilon}$ with increasing field and
thus, one may argue that the small exponents we observe for all
U/I transitions are related to tricritical behavior.
However, above $\rm H_c$ the range of small
$\rm 2\beta_{\epsilon}$ is much larger
than below $\rm H_c$ (Fig.~\ref{critis}).
This pronounced difference indicates a critical behavior
within the I phase, which differs from the rather
usual one found at the U/D transitions.
In principle the small values of $\rm 2\beta_{\epsilon}$
might also arise from a different strain
order parameter coupling in the I phase.
However, as we will show below, the temperature dependence of the
spontaneous strain for high fields indicates
the usual linear quadratic coupling for the U/I transitions, too.
Irrespective of whether the critical behavior at the U/I transition is
intrinsically characterized by a smaller $\beta_{\epsilon}$ or the
tricritical behavior is found in a larger field range,
the critical exponents indicate a qualitative difference
between the D and I phases.

\subsection{Magnetic field dependence of the spontaneous strain}
\label{SUBBY3}

From our data of the thermal expansion coefficient in magnetic fields
we can follow the spontaneous strain as a function of magnetic field
as well as a function of temperature.
Fig.~\ref{3D} presents the development of the spontaneous
strain as a function of H and T. In addition, we show
the spontaneous strain as a function of temperature
for some representative fields in Fig.~\ref{strains}.
The magnetic field dependence of $\rm \epsilon$ is dominated by the field
induced phase transitions.
At low temperatures there is only a moderate decrease of $\rm \epsilon$
with increasing field within the D phase, i.e. for $\rm H \lsim 12T$.
At the discontinuous D/I transition a strong decrease
of $\rm \epsilon$ occurs in a narrow field range followed by
a weaker decrease within the I phase. Close to the
D/I boundary the temperature dependence of $\rm \epsilon$ reflects
the competition between these phases leading to multiple transitions.

The field dependence of $\epsilon$ at
fixed temperatures is closely related to the field dependence of the
lattice constant, i.e. to our measurements of the
magnetostriction presented in section~\ref{RESULTS}.
However, small but significant differences occur,
when comparing the field dependence of the lattice constant $a$
quantitatively with $\rm\epsilon$.
The field dependence of $\rm \epsilon$ is always smaller than the
magnetostriction at the same temperature (see Figs.~\ref{ms}
and~\ref{strains}). These differences arise from the finite
magnetostriction above $\rm T_{SP}$. Whereas
the spontaneous strain is zero and thus field independent
above $\rm T_{SP}$, the lattice constant does depend on the magnetic field.
From the definition of the spontaneous strain one obtains:
\begin{eqnarray}
\label{correct}
\rm
\epsilon(H,T)-\epsilon(O,T)    & = &
\int\limits_{T_0>T_{SP}}^T (\delta\alpha(H,T') dT' -
\delta\alpha(0,T')) dT' =
\frac{\Delta L(H,T)}{L} - \frac{\Delta L(H,T_0)}{L}
\end{eqnarray}
Eq.~\ref{correct} means that the field dependence of
$\rm \epsilon$ at a temperature T is given by the difference of
the magnetostriction at the same temperature and the magnetostriction at
a temperature $\rm T_0$ above $\rm T_{SP}$.
Subtracting the magnetostriction data at $\rm T_0=20 K$ from the low
temperature magnetostriction leads to a good agreement between
the spontaneous strain derived from
the magnetostriction and thermal expansion data, respectively.
That means that the lattice constant and
the spontaneous strain follow variations of T and H
in a reversible way in almost the entire field and temperature range.
The only exception is a small region around the D/I phase boundary
at low temperatures where hysteresis effects are present in the
magnetostriction.

Please note that the need for correcting the magnetostriction
data before comparing with $\epsilon$ does not imply
that the magnetostrictive effects present in the U phase are also
relevant in the dimerized phase. The correction
according to Eq.\ref{correct} is just a consequence of the
definition of the spontaneous strain. There is, however, some
experimental evidence for an unexpected magnetoelastic coupling
in the D phase. A closer inspection of the magnetostriction data
in Fig.~\ref{ms} below about 5K shows a small increase of
$a$ with increasing field in the D phase. Since
the sign of this effect differs from the magnetostriction found
at the field driven transitions, it is probably
not related to the field dependence of the spontaneous strain.
However, there seems to be a correlation
to the magnetostriction in the U phase, which shows the same sign and
order of magnitude. At present we have no explanation
of this finite magnetostriction in the dimerized phase at low
temperatures. Detailed investigations of the other lattice constants
and also for crystals with different impurity concentrations
are in progress to clarify this low temperature behavior.

A detailed discussion of the temperature dependence
of $\rm Q$ in external fields will be given in the following
section. Here we will restrict the discussion of $\epsilon(H)$
to the low temperature range, i.e. for $\rm T \rightarrow 0$.
Within the D phase the scaling between the spin--Peierls
order parameter and the spontaneous strain is valid as discussed in
section~\ref{ZERO}.
Thus, it is apparent from Fig.~\ref{strains}
that the field dependence of the order
parameter at low temperatures is extremely weak.
Extrapolating the data to $\rm T = 0$ our measurements are
consistent with a field independent $\rm Q^2(0)$. In other words,
not only the wave vector of the distortion is pinned and thus field
independent up to the critical field $\rm H_{D/I}$ but also
the amount of the structural distortion for $\rm T \rightarrow 0$
does not change as a function of H. This is in striking
contrast to the strong decrease of the energy gap in the magnetic
excitations due to the Zeeman splitting of the triplet
state~\cite{fujita95a}.
The scaling between structural deformation $\rm Q$ and the energy gap,
theoretically expected and observed in $\rm CuGeO_3$ for zero magnetic
field~\cite{harris94,fujita96a}, is obviously not present at finite fields.
However, we emphasize that this is not in contrast to theoretical treatments
of the spin--Peierls transition. For example, calculations
within the XY model really yield a field independent
order parameter at T = 0~\cite{mueha}.

A similar simple interpretation of $\rm \epsilon (H)$
as a measure of the amount of the structural distortion
is not possible at and above the D/I phase transition.
The spontaneous strain for an individual field is a consequence
of the strain order parameter coupling, too. According to our data
$\rm \epsilon$ is reduced but still large for the U/I transitions.
However, in contrast to the behavior at low fields
the character of the lattice distortion now changes as a function
of H. This has been studied with x--ray diffraction showing
a field dependent wave vector $q$ of the lattice distortion at and
slightly above the D/I transition~\cite{kiryukhin95b}.
In this field range it is impossible to correlate the
field dependence of the spontaneous strain to a single quantity.
Assuming a homogeneous incommensurate modulation of the lattice
the decrease of the spontaneous strain may be a consequence of
a reduced amplitude of the distortion and/or a consequence of a wave vector
dependent strain order parameter coupling.
Similarly, assuming a domain--wall picture,
the decrease of $\rm \epsilon$ may be a consequence of a reduced amplitude
of the distortion within the individual domains and/or the number
of domain walls. With the present knowledge
of the structural deformation in the I phase it seems neither possible
to favor one of the two models nor to relate the $\rm \epsilon(H)$
to a structural characteristic of the I phase.
However, note that the large anomalies
at the D/I transition clearly show structural differences
between the D and I phases. Moreover, as visible in
Fig.~\ref{strains} and also in the magnetostriction data,
there is no indication for
a field independent $\rm \epsilon$ for $T\rightarrow 0$ as we observe
within the D phase. Obviously the structural parameter(s) determining
the spontaneous strain in the I phase continuously change with
increasing field.

\subsection{Temperature dependence of the spin--Peierls
order parameter in external fields}

The field dependencies of the quantities discussed in the last three
subsections, $\rm \Delta\alpha(T_{SP})$, 2$\beta_{\epsilon}$, and
$\rm \epsilon (T=const.)$, mainly reflect the field driven phase transitions.
In particular, there are only minor changes within
the D phase.
A completely different picture is obtained when considering
the temperature dependence of the order parameter.
Within the D phase the temperature dependence of $\rm \epsilon$ well
below $\rm T_{SP}$ systematically
increases with increasing field (see Fig.~\ref{strains}).
For fields of 15 and 16 Tesla, i.e. in the I phase, a
much smaller slope is obtained. In order to analyse the
temperature dependencies of the (squared) order parameter we will not
consider the spontaneous strains plotted in Figs.~\ref{3D} and \ref{strains}.
Instead of that we will directly investigate their temperature derivatives
$\delta\alpha$, where small changes of the temperature dependence
of $\rm Q^2$ show up more clearly.

As a starting point of this discussion we investigate
the field dependence of $\rm \delta\alpha$ in the low temperature range.
Fig.~\ref{at5K}(a) shows $\delta\alpha$ at a fixed temperature of 5K
as a function of magnetic field. It is apparent
that this field dependence strongly differs from
those discussed in the last sections. In particular, it
strongly differs from the field dependence of the anomaly
size $\Delta\alpha$ at $\rm T_{SP}$.
Within the D--phase $|\delta\alpha|$ systematically increases with H.
At the D/I boundary $\rm |\delta\alpha(5K)|$
jumps back to smaller values and for
higher fields it is roughly constant.
We emphasize that this strong field dependence of
$\rm \delta\alpha(5K)$ in
the D phase is not due to the decreasing $\rm T_{SP}(H)$.
As displayed in Fig.~\ref{at5K}(b) at a fixed reduced temperature
($\rm T_{SP}/2$) a very similar field dependence of
$\rm \delta\alpha$ is present in the D phase. However, in this
representation $|\delta\alpha|$ is significantly
smaller in the I phase than in the D phase. This difference can be traced
back to the different sizes of the anomalies at $\rm T_{SP}$
(see Fig.~{\ref{peaks}). In Fig.~\ref{at5K}(c) the different sizes are taken
into account by
considering relative values, i.e. $\delta\alpha(T_{SP}/2)~/~\Delta\alpha$,
as a function of H.
The result is rather surprising. The zero field value
roughly coincides with the values at very high magnetic fields,
i.e. with those of the I phase. However, the pronounced field dependence of
$\rm \delta\alpha$ within the D is still present.

In the following we first analyze $\rm Q^2(T)$
for different magnetic fields within the D phase.
In order to avoid complications due to
tricritical behavior and the additional anomalies in the vicinity of the
D/I phase boundary (see Fig.~\ref{al1213}) we restrict the discussion
to fields below 11 Tesla.
From the data analysis given in the previous sections
one concludes that the decrease of the transition
temperature is the main consequence of the magnetic field
in this field range, whereas only minor changes occur in all
structural quantities. In particular,
the zero temperature value of $\rm Q^2$ as well as its critical behavior
are roughly field independent. Thus, one might
even expect a universal relationship
between the order parameter and the reduced temperature
$\rm t=(T_{SP}(H)-T)~/~T_{SP}(H)$
within the D phase, i.e. a simple scaling behavior
$\rm Q(t,H) = Q(t,0)$.

However, this universality is not present at all in the D phase.
To demonstrate this we consider $\rm \delta \alpha \cdot T_{SP}(H)$, i.e.
the derivative
$\rm \partial \epsilon(t) / \partial t \propto \partial Q^2(t) / \partial t$
(Fig.~\ref{scal1}).
Obviously there are pronounced
systematic differences of the temperature dependence of Q in
different magnetic fields.
Close to the transition temperature $\rm \partial Q^2(t) / \partial t$
decreases with H, whereas at low temperatures the opposite
field dependence is present. At
$\rm t \simeq 0.44$ all curves meet in a single point,
i.e. $\rm \partial Q^2(t)/ \partial t$
is field independent for this particular reduced temperature.
It is not possible to improve the agreement between
the curves for different magnetic fields
in the whole temperature range by any normalization of
$\rm \delta \alpha$ (see inset of Fig.~\ref{scal1}).
For example, when investigating $\rm Q^2(t)$ divided
by its zero temperature value, i.e. the quantity
shown for the zero field data in Fig.~\ref{bcsop},
a systematic increase of the derivative
with H is obtained at low temperatures.
Thus, the BCS--like low temperature behavior of Q(T)
present for H = 0 (see Fig.~\ref{bcsop})
does significantly change with increasing magnetic field.

Instead of a field independent Q(t) we empirically find
a -- to our opinion -- rather surprising result.
The temperature dependence of the order parameter
in different magnetic fields is universal on an {\em absolute}
temperature scale. Fig.~\ref{scal2} shows a plot
of $\rm \delta \alpha/\sqrt{T_{SP}(H)}$
versus $\rm (T-T_{SP}(H))$:
All the curves from 0 up to 11 Tesla, i.e. in the
entire field range of the dimerized phase,
perfectly agree within the experimental resolution.
Note that the factor $\rm 1/\sqrt{T_{SP}(H)}$, which is used
to scale the y--axis, is necessary
to take into account the slight decrease
of the anomaly size $\rm \Delta\alpha$.
We emphasize that the temperature axis for the different
curves in Fig.~\ref{scal2} is not scaled but simply
shifted by the field dependent transition temperature.
To our knowledge there is no theoretical calculation
of the temperature dependence of the spin--Peierls order parameter
in magnetic fields, at least none that is consistent with our findings.
For example from the exact solution of the XY--model one obtains
indeed an increase of $\rm \partial Q^2 / \partial t$ with H
at low temperatures. However, the opposite field dependencies
for small and large reduced temperatures as well
as the crossing at $\rm t \simeq 0.44$ are not obtained~\cite{mueha}.

Now we turn to the temperature dependence of the spontaneous strain
in the I phase, which compares
well with that in zero magnetic field. At first sight any similarity
between high field and zero field data is covered by the
much smaller size of the anomalies at the U/I transition.
However, the data in Fig.~\ref{at5K} already give a first
hint on this similarity: After normalizing the
$\delta\alpha$ axis the temperature derivatives of the spontaneous
strains at high fields nearly agree with that
in zero field at $\rm T_{SP}/2$ (Fig.~\ref{at5K}c).
This holds at $\rm T_{SP}/2$ but also in the entire low temperature
range. On reduced scales the temperature dependencies
of $\delta\alpha$ in H = 0 and H = 16T are very similar
(see Fig.~\ref{scal3}).
Only close to $\rm T_{SP}$ slight differences
are present, which have already been discussed
in connection with the critical exponent $\rm 2\beta_{\epsilon}$
(see Fig.~\ref{critis}).
The scaling used to obtain agreement between
H= 0 and H = 16 T data is quite natural, since the first
-- considering the reduced temperature $\rm T_{SP}/2$ -- takes
into account the smaller $\rm T_{SP}$ and the second
the reduced anomaly size $\rm \Delta\alpha$ and/or the
reduced spontaneous strain in the I phase (see
Figs.~\ref{peaks} and~\ref{strains}).
From Fig.~\ref{scal3} one thus concludes
that apart from the absolute values of both $\rm \epsilon(T=0)$
and $\rm T_{SP}$ the spontaneous strain
in the incommensurate phase corresponds to that
found in the D phase at H = 0. This similarity is also found
for the other fields with U/I transitions.

To connect $\rm \epsilon$ in the I phase
to the order parameter of the incommensurate
modulation, we can follow the procedure applied for H = 0.
A strain order parameter coupling causes a spontaneous strain
and the temperature dependence of $\epsilon$
measures that of the incommensurate modulation (Eq.~\ref{landi}
and~\ref{scali}).
There are no neutron scattering measurements of the superstructure
reflections. Thus, we can not prove a linear
quadratic coupling, as we did for smaller fields in Fig.~\ref{epsop}.
However, the similarity between
zero and high magnetic fields strongly indicates
the same linear quadratic
strain order parameter coupling in the I phase.
This means that the spontaneous strain yields a measure of the squared
order parameter of the I phase, too.
In particular, at low temperatures the order
parameter of the I phase follows roughly a BCS mean
field behavior (see Fig.~\ref{bcsop}).
Up to now  there are -- to our knowledge -- neither measurements
nor calculations of $\rm Q(T)$ in the I phase of any spin--Peierls compound.
Thus, at present we can neither compare our result to other findings
nor judge whether the
presented data give a possibility to distinguish between
the different models
of the structural distortion in the I phase, i.e. domain walls
or sinusoidal modulation.

\section{Conclusions}

Using a high resolution capacitance dilatometer we
have investigated structural and thermodynamic properties
of the inorganic spin--Peierls compound $\rm CuGeO_3$.
The temperature and field dependence of the lattice constant $a$ have
been studied via measurements of the thermal expansion and the
magnetostricton, respectively.
Pronounced anomalies are found at all phase transitions
present in the characteristic field temperature phase diagram of
spin--Peierls compounds. Thus, clear structural
differences between the three phases are established from our data.
The very large anomalies we observe at all phase transitions
allow for a precise determination of the H--T phase
diagram of $\rm CuGeO_3$.
Moreover, our data show that
all phase boundaries strongly depend on pressure,
since the anomalies of the thermal expansion and
the magnetostriction are related
to the uniaxial pressure dependencies of the transition temperature
and field, respectively.
Summarizing our detailed comparison of the experimental H--T phase diagram
to existing theories, one concludes that there are pronounced discrepancies
to the soliton theory. On the other
hand the predictions given by
Cross as well as by Bulaevskii et al. are in rough agreement with the
experimental results. This holds for the first order character of the D/I
transition and also for a quantitative comparison of
the transition fields at the phase boundaries.
However, clear deviations from the latter theories are also present.
The temperature at the tricritical point calculated by Bulaevskii et al.,
for instance,
is significantly too small. With respect to this point
the Cross theory agrees
very well with the data. Moreover, the overall shape of the U/D phase boundary
follows Cross' calculations although the field scales differ slightly.
The ''effective'' field acting in $\rm CuGeO_3$ is about 10\%
larger than predicted by theory. This enhancement indicates that particular
magnetic properties of $\rm CuGeO_3$ determine the
exact positions of the phase boundaries.

Our high resolution measurements of the lattice constant do not only
yield the phase boundaries. Properties
of the ordered phase can also be extracted. From a simple treatment
within Landau theory one obtains a scaling between
the squared spin--Peierls order parameter and the spontaneous strain,
which is experimentally confirmed for the U/D transitions.
Since the anomalous contribution of the thermal expansion
corresponds to the temperature derivative of the spontaneous strain,
our data yield a high resolution measurement of the field and temperature
dependence of the order parameter.
The temperature dependence of the order parameter
is strongly affected by fluctuations.
In zero magnetic field, for example, there are pronounced
deviations from a mean field behavior in the entire
temperature range. Close to $\rm T_{SP}$
a power law as a function of the reduced temperature is
found and the extracted exponent is consistent
with the predictions for a 3d--Ising transition.
In order to describe $\rm Q^2(T)$
at low temperatures by a BCS--like behavior one
has to assume a 10\% reduction of $\rm T_{SP}$, which is due
to fluctuations.

For the discussion of the field dependence of the spontaneous
strain one has to discriminate between the different low temperature phases.
Up to 12 Tesla $\rm \epsilon(T,H)$ measures the dimerization,
i.e. the order parameter of the D phase. There are
three main conclusions from our data in this field range.
At very low temperatures the order parameter is
nearly field independent.
The critical behavior at the U/D transition is also rather
field independent for a wide field range.
The third finding is rather surprising.
The temperature dependence of $\rm Q^2$
is universal on an absolute temperature scale,
i.e. when plotting $\rm Q^2$ versus $\rm T_{SP}(H)-T$.
In contrast pronounced systematic
differences are present after scaling the
temperature axes, i.e. when analyzing $\rm Q^2$
as a function of the reduced temperature
$\rm {T_{SP}(H)-T \over T_{SP}(H)}$. Thus,
the similarity to a BCS--like behavior found for H = 0
rapidly vanishes with increasing field.

The interpretation of our findings involving the incommensurate
phase is more complicated. In this field range
the wave vector of the structural deformation changes with H
and thus the spontaneous strain is not related to a single, i.e.
field independent, structural parameter.
The field dependence of
$\epsilon$ might be a consequence of a field dependent
order parameter as well as a field dependent strain order parameter
coupling. Nevertheless, the temperature dependence of
$\epsilon$ at a given field reflects that of the corresponding order parameter.
At a fixed temperature, e.g. for $\rm T\rightarrow 0$,
there is a strong decrease of $\epsilon$ with increasing field.
This decrease is most pronounced at the first order D/I transition
itself, where $\epsilon$ reduces to about 40\% of its zero field
value. Within the I phase $\epsilon$ shows a further decrease with H, which
indicates continuous structural changes as a function
of field.

For fixed fields we analyzed the critical behavior
of $\epsilon$ close to $\rm T_{SP}$ as well as its
temperature dependence well below the transition.
The critical exponents in the I phase are significantly smaller than those
found in the D phase. Although we could investigate the I phase only
in a rather small field range, the critical exponents indicate a
qualitative difference between U/D and U/I transitions.
Apart from the region close to $\rm T_{SP}$
the temperature dependence of $\rm \epsilon$ in the I phase
compares well with that of the squared order parameter for H = 0.
Thus, the incommensurate lattice modulation roughly follows
a BCS mean field behavior for $\rm T \longrightarrow 0$,
similar to the dimerization in zero magnetic field.

{\bf Acknowledgements}

We are grateful to W. Brenig, E. M\"uller--Hartmann, and A. Kl\"umper for
stimulating discussion on many aspects of this work. We thank
H. Micklitz and B. Sullewsky for critical reading of the manuscript.
A.R. and G.D. acknowledge NEDO for financial support. U.A.
acknowledges support by the Graduiertenkolleg GRK14 of the
Deutsche Forschungsgemeinschaft. This work was supported
by the Deutsche  Forschungsgemeinschaft through SFB 341.

\begin{figure}       
\caption[]{Thermal expansion of $\rm CuGeO_3$
along the $a$--axis in different magnetic fields.
The solid line represents the extrapolated
low temperature behavior $\rm \alpha_{extr.}$ of the U phase.}
\label{alueber}
\end{figure}

\begin{figure}       
\caption[]{Thermal expansion of $\rm CuGeO_3$
along the $a$--axis in different magnetic fields given in the
figure. The curves are shifted by $\rm 5 \cdot 10^{-6}/K$ for clarity.}
\label{algenau}
\end{figure}

\begin{figure}        
\caption[]{Thermal expansion
of $\rm CuGeO_3$ along the $a$--axis
in magnetic fields close to the D/I phase boundary.
The additional anomalies reflect temperature dependent
transitions between the D and I phases and vice versa.}
\label{al1213}
\end{figure}

\begin{figure}         
\caption[]{Magnetostriction of $\rm CuGeO_3$
along the $a$--axis at different temperatures given in the figure.
Upper panel: Discontinuous transitions between D and I phases.
With increasing T (indicated by the arrows) the character of the transition
gradually changes from rather strong to weak first order. \\
Inset: Field derivatives $\rm 1/L\cdot \partial L / \partial H$ at 3.4K
measured with increasing ($\circ$) and
decreasing ($\bullet$) magnetic field.
The hysteresis, resolved up to $\rm \sim 11K$,
systematically decreases with decreasing T.             \\
Lower panel: Continuous transitions between D and U phases.
The transition fields as well as the overall magnetostriction
rapidly decreases with increasing T.
Within the U phase (T=18K and T = 14K for H $\gsim 8$T) a positive
magnetostriction is present.}
\label{ms}
\end{figure}

\begin{figure}          
\caption[]{Magnetic field temperature phase diagram of $\rm CuGeO_3$
derived from thermal expansion (circles)
and magnetostriction (triangles) data.  The solid line shows the result of
Cross' calculation. An almost perfect agreement to the experimental
results for the entire U/D phase boundary
is obtained by dividing the calculated field scale by 1.12
(dotted line, see text).  \\
Inset: Hysteresis at the D/I phase boundary obtained from the
magnetostriction measured with increasing
(closed triangles) and decreasing field (open triangles).}
\label{btphadi}
\end{figure}

\begin{figure}          
\caption[]{Comparison between the spontaneous strain ($\bullet$)
and the intensity of a superstructure reflection ($\circ$) found from
neutron diffraction~\cite{regnault96a}. Main part: Zero field data.
Inset: Spontaneous strain in $\rm H=10T$ and neutron scattering
intensity in $\rm H=9.85 T$.}
\label{epsop}
\end{figure}

\begin{figure}          
\caption[]{Symbols: Temperature dependence of the
squared order parameter in zero magnetic field as obtained
from the thermal expansion.\\
Solid line: Power law ($\rm (14.35-T)^{0.61}$
representing the order parameter fluctuations close to $\rm T_{SP}$.\\
Dotted line: BCS mean field behavior for $\rm T_{SP} = 14.35K$. \\
Dashed line: BCS mean field behavior for a hypothetical
$\rm T^{mf}_{SP} = 15.8K$ (see text).}
\label{bcsop}
\end{figure}

\begin{figure}          
\caption[]{Left panel: Anomaly size
$\rm \Delta\alpha \equiv max(\alpha-\alpha_{extr})$
as a function of magnetic field. \\
Right panel: $\rm \Delta\alpha / \sqrt{T_{SP}}$ versus magnetic field
(see text).\\
The dashed vertical lines denote the field at the tricritical point.}
\label{peaks}
\end{figure}

\begin{figure}         
\caption[]{Critical behavior of $\rm \epsilon$.
Upper panel: $\rm \epsilon$ versus reduced temperature
on a double logarithmic scale. The curves are shifted for clarity.\\
Lower panel: Magnetic field dependence of the critical
exponents $\rm 2\beta_{\epsilon}$.}
\label{critis}
\end{figure}

\begin{figure}          
\caption[]{Spontaneous strain ($\circ$) as a function of temperature and
magnetic field. The phase boundaries between the U, D, and I phases
are given by the closed symbols ($\bullet$).}
\label{3D}
\end{figure}

\begin{figure}         
\caption[]{Spontaneous strain as a function of T for
several magnetic fields given in the figure.
The different field ranges represent the
different kinds of low temperature phases.
At H=12.5 T the competition between D and I phase close to the
D/I phase boundary is seen.}
\label{strains}
\end{figure}

\begin{figure}         
\caption[]{Different representations of the temperature derivative
$\rm \delta\alpha =\partial \epsilon/\partial T$
$\rm \propto \partial Q^2/\partial t$
of the spontaneous strain versus magnetic field: \\
a.): $\rm \delta\alpha(H)$ at a fixed temperature $\rm T=5K$.  \\
b.): $\rm \delta\alpha(H)$ at a fixed reduced temperature
     $\rm t= T_{SP}(H)/2$\\
c.): $\rm -\delta\alpha / \Delta\alpha$  at
     $\rm t= T_{SP}(H) /2$ (see text).             }
\label{at5K}
\end{figure}

\begin{figure}         
\caption[]{Derivative 
$ \delta\alpha \cdot T_{SP}(H) \propto \rm {\partial Q^2(t) \over \partial t}$
versus reduced temperatures t for different magnetic fields
given in the figure.
Inset: Same data after an arbitrary normalization of the y--axis.
}
\label{scal1}
\end{figure}

\begin{figure}         
\caption[]{Universal temperature dependence of the D phase order parameter.
Note, that the temperature axis is only
shifted but not scaled by $\rm T_{SP}(H)$.
The division of
$\rm \delta\alpha$ by $\rm \sqrt{T_{SP}(H)}$ takes
into account the slight decrease of the anomaly size with H (see text).}
\label{scal2}
\end{figure}

\begin{figure}         
\caption[]{Comparison between the derivatives
$ \delta\alpha \cdot T_{SP}(H) \propto \rm {\partial Q^2(t) \over \partial t}$
versus reduced
temperature t in $\rm H=0$ ($\circ$; left y--scale) and
$\rm H=16 T$ ($\bullet$; right y--scale).}
\label{scal3}
\end{figure}


\begin{thebibliography}{10}
\bibitem{otherdir}
The measurements of the thermal expansion coefficients along the $b$-- and
  $c$--axis, respectively, will be presented in a forthcoming publication.
\bibitem{mehrkommt}
A detailed discussion of the results of the magnetostriction measurements along
  the different crystal axes will be given in a forthcoming publication.
\end{thebibliography}
\end{document}